\documentclass[conference,letterpaper]{IEEEtran}

\makeatletter

\IEEEoverridecommandlockouts
\usepackage{cite}
\usepackage{amsmath,amssymb,amsfonts}
\usepackage{algorithm}
\usepackage{algpseudocode} 

\usepackage{graphicx}
\usepackage{textcomp}
\usepackage{xcolor}
\usepackage[none]{hyphenat}
\usepackage{float}
\usepackage{url}
\usepackage{vcell}

\definecolor{rotcol}{RGB}{230,242,255}  
\definecolor{entcol}{RGB}{240,240,240}  
\definecolor{linkcol}{RGB}{255,243,205} 
\definecolor{gateborder}{RGB}{120,120,120} 

\newcommand{\gatebox}[2]{%
  \gate{\fcolorbox{gateborder}{#1}{\hspace{1.5pt}$#2$\hspace{1.5pt}}}%
}
\newcommand{\rotgate}[1]{\gatebox{rotcol}{#1}}
\newcommand{\entgate}[1]{\gatebox{entcol}{#1}}
\newcommand{\linkgate}[1]{\gatebox{linkcol}{#1}}

\usepackage{balance} 

\usepackage[utf8]{inputenc}
\usepackage[T1]{fontenc}
\usepackage{pdflscape}

\usepackage{array}   
\usepackage{booktabs} 

\usepackage{siunitx}
\usepackage{numprint}
\usepackage{tabularx}
\usepackage{xltabular}
\usepackage{makecell}
\usepackage{multirow}
\usepackage{adjustbox}
\usepackage{caption}
\usepackage{subcaption}
\usepackage{mweights}
\usepackage{url}
\usepackage{cleveref}
\usepackage{bchart}

\usepackage{stfloats}  

\usepackage{tikz}
\usetikzlibrary{matrix, positioning, arrows.meta, calc}
\usepackage{quantikz}
\usepackage{pgfplots}
\usepackage{pgfplotstable}
\pgfplotsset{compat=1.18}

\def\BibTeX{{\rm B\kern-.05em{\sc i\kern-.025em b}\kern-.08em
    T\kern-.1667em\lower.7ex\hbox{E}\kern-.125emX}}

\usepackage{eso-pic}
\usepackage{geometry}
\begin{document}

\title{

\vspace*{1mm} Quantum Kernels for Audio Deepfake Detection Using Spectrogram Patch Features}

\author{
    \IEEEauthorblockN{Lisan Al Amin\IEEEauthorrefmark{1},
                      Rakib Hossain\IEEEauthorrefmark{1},
                      Mahbubul Islam\IEEEauthorrefmark{2},
                      Faisal Quader\IEEEauthorrefmark{3},
                      and Thanh Thi Nguyen\IEEEauthorrefmark{4}\IEEEauthorrefmark{5}}
    
    \IEEEauthorblockA{\IEEEauthorrefmark{1}Potomac Quantum, USA}
    \IEEEauthorblockA{\IEEEauthorrefmark{2}United International University, Dhaka, Bangladesh}
    \IEEEauthorblockA{\IEEEauthorrefmark{3}University of Maryland, College Park, MD, USA}
    \IEEEauthorblockA{\IEEEauthorrefmark{4}Monash University, Melbourne, Victoria, Australia}
    \IEEEauthorblockA{\IEEEauthorrefmark{5}University of the Sunshine Coast, Queensland, Australia}
    \IEEEauthorblockA{E-mails: lisanalamin@gmail.com, rakib.sat18@gmail.com, mahbub120.eee@gmail.com, \\
                      fquader1@umbc.edu, tnguyen5@usc.edu.au}
}

\maketitle

\begin{abstract}
Quantum machine learning has emerged as a promising tool for pattern recognition, yet many audio-focused approaches still treat spectrograms as generic images and do not explicitly exploit their time-frequency structure. We propose Q-Patch, a quantum feature map tailored to audio that encodes local time-frequency patches from mel-spectrograms into quantum states using shallow, hardware-efficient circuits with adjacency-aware entanglement. Each selected patch is summarized by a compact four-dimensional acoustic descriptor and mapped to a four-qubit circuit with depth at most three, enabling practical quantum kernel construction under near-term constraints. We evaluate Q-Patch on an audio spoofing detection task using a controlled, balanced protocol and compare it with size-matched classical baselines. Q-Patch improves discrimination between bona fide and spoofed samples, achieving an area under the receiver operating characteristic curve (AUROC) of 0.87, compared with 0.82 for a radial basis function support vector machine (RBF-SVM) trained on the same patch-level features. Kernel-space analysis further reveals a clear class structure, with cross-class similarity around 0.615 and within-class self-similarity of 1.00. Overall, Q-Patch provides a practical framework for incorporating time-frequency-aware representations into quantum kernel learning for audio authenticity assessment in low-resource settings.
\end{abstract}


\section{Introduction}
Recent advances in generative speech have made modern text-to-speech (TTS) and voice conversion (VC) systems capable of producing audio that is often difficult to distinguish from genuine human speech. While these technologies support accessibility and content creation, they also introduce serious security risks, including impersonation, fraud, and misinformation. As a result, reliable audio spoofing detection has become an important research problem \cite{_1}.

Benchmarks such as ASVspoof 2019 \cite{_2} and ADD 2022~\cite{_3} have accelerated progress, but many detectors still struggle under unseen attacks and real-world distortions, including channel effects, background noise, compression, and replay artifacts \cite{_4,_5}. In addition, many existing approaches process spectrograms as generic images, which can overlook important time--frequency structure in speech.

Quantum machine learning offers an alternative framework for representation learning and similarity estimation through quantum state overlaps \cite{_8}. In particular, quantum kernel methods are attractive in limited-data settings, yet their use in speech security remains limited, and most existing pipelines are not designed around the structure of time--frequency audio representations \cite{_9}.

Motivated by these gaps, we propose \textit{Q-Patch}, a quantum feature-mapping framework for audio spoofing detection that encodes informative local time--frequency patches into shallow quantum circuits. The main contributions of this work are as follows:
\begin{itemize}
\item We introduce a patch-based audio representation that summarizes local time--frequency regions using compact and interpretable acoustic descriptors.
\item We design a shallow quantum feature map with adjacency-aware entanglement, limiting circuit depth to three layers and qubit usage to 4--8 qubits for compatibility with Noisy Intermediate-Scale Quantum (NISQ) constraints.
\item We evaluate the proposed framework in a controlled low-resource setting against size-matched classical baselines and analyze the induced kernel space using both classification metrics and similarity structure.
\end{itemize}

We validate Q-Patch on controlled audio data derived from LJ Speech \cite{_17}. The results suggest that time--frequency-aware quantum feature maps can provide a useful inductive bias for audio authenticity discrimination while maintaining a shallow circuit design suitable for near-term quantum settings.

The rest of the paper is organized as follows. Section~\ref{sec:rw} reviews related work, Section~\ref{sec:metd} presents the proposed methodology and experimental protocol, whilst Section~\ref{sec:results} reports the results and discussion, and Section~\ref{sec:conclusion} concludes the paper.

\section{Related Work}
\label{sec:rw}

This section reviews prior work in audio deepfake detection and quantum kernel learning, and then highlights the gap addressed by \textit{Q-Patch}.

\subsection{Audio Deepfake Detection and Anti-spoofing}
The increasing realism of TTS and VC systems has driven substantial progress in audio deepfake detection. Surveys by Khanjani \textit{et al.} \cite{_1,_10} and Pham \textit{et al.} \cite{_9} trace the field from handcrafted features to deep learning, supported by benchmarks such as ASVspoof 2019 \cite{_2} and ADD 2022 \cite{_3}. These benchmarks provide standardized protocols and include degraded and channel-distorted conditions that better reflect deployment scenarios.

Current state-of-the-art systems are dominated by deep neural architectures. RawNet2 \cite{_7} and DeepLASD \cite{_8}, for example, learn discriminative representations directly from waveforms or spectrogram-like inputs. Related evidence also suggests that compact models can remain effective under strong resource constraints; Al Amin \textit{et al.} \cite{_18} show that sparse subnetworks can retain strong detection capability. Related multimodal studies have also shown that self-supervised vision and speech encoders can provide effective compact representations in applied human-centered settings \cite{hossain2025risk}, reinforcing the broader case for structured feature design beyond large end-to-end architectures. Despite these advances, robustness remains a key challenge, with performance often degrading under unseen attacks, distribution shifts, and realistic distortions \cite{_4,_5}. This motivates approaches that improve generalization and better exploit time--frequency structure.

\subsection{Quantum Machine Learning and Kernel Methods}
Quantum machine learning investigates whether quantum feature representations and quantum-native similarity measures can improve learning performance. In the context of kernel methods, Schnabel and Roth \cite{_6} show that feature-map design plays a central role in kernel expressivity. Egginger \textit{et al.} \cite{_11} and Innan \textit{et al.} \cite{_13} further emphasize the importance of circuit depth, encoding strategy, and trainability in quantum kernel quality.

Applications of quantum kernels are beginning to emerge across domains. Beaulieu \textit{et al.} \cite{_12} apply them to manufacturing defect detection, while Tran \textit{et al.} \cite{_15} explore their use in biomedical speech analysis. Broader surveys \cite{_14,_16} also stress the importance of shallow, hardware-aware designs in the Noisy Intermediate-Scale Quantum era. However, quantum-kernel methods remain underexplored for audio deepfake detection, especially when the representation is designed explicitly around spectrogram time--frequency structure.

\subsection{Gaps and Motivation}
The literature points to three main gaps. First, robustness to unseen attacks and real-world distortions remains limited \cite{_4,_5}. Second, low-resource detection settings are still less studied, despite their practical relevance when new spoofing methods appear. Third, the use of quantum kernels for audio deepfake detection remains limited, and existing quantum machine learning approaches rarely incorporate inductive biases tailored to spectrogram geometry\cite{al2025reliable}. These gaps motivate \textit{Q-Patch}, a time--frequency-aware quantum kernel framework designed for structured spectrogram patches, limited-data discrimination, and compatibility with near-term quantum hardware.

\section{Methodology}
\label{sec:metd}

This section describes the proposed Q-Patch framework, including data preparation, patch-level feature construction, quantum embedding, training procedure, baseline design, and implementation details.

\subsection{Overview}
We introduce \textit{Q-Patch}, a quantum feature-mapping framework designed to capture local time--frequency structure in speech spectrograms using shallow entangling circuits. The method follows three stages: (i) computing an utterance-level log-mel representation, (ii) partitioning the time--frequency plane into small patches and summarizing each patch with a compact vector of interpretable statistics, and (iii) embedding the selected patch summaries into quantum states to form a fidelity-based kernel for a Quantum Support Vector Machine (QSVM) \cite{_11,_12}.

To isolate the contribution of the proposed quantum feature map, we compare Q-Patch with matched classical baselines, including a radial basis function support vector machine (RBF-SVM) trained on the same patch summaries and a compact convolutional neural network (CNN) with $\leq$100k parameters trained directly on spectrograms. Algorithm~\ref{alg:qpatch_extract}, presented later in this section, specifies preprocessing, patching, and feature construction, whereas Algorithm~\ref{alg:qsvm_kernel} details quantum-kernel computation and QSVM training/inference.

\begin{figure*}[t]
    \centering
    \includegraphics[width=1.0\linewidth]{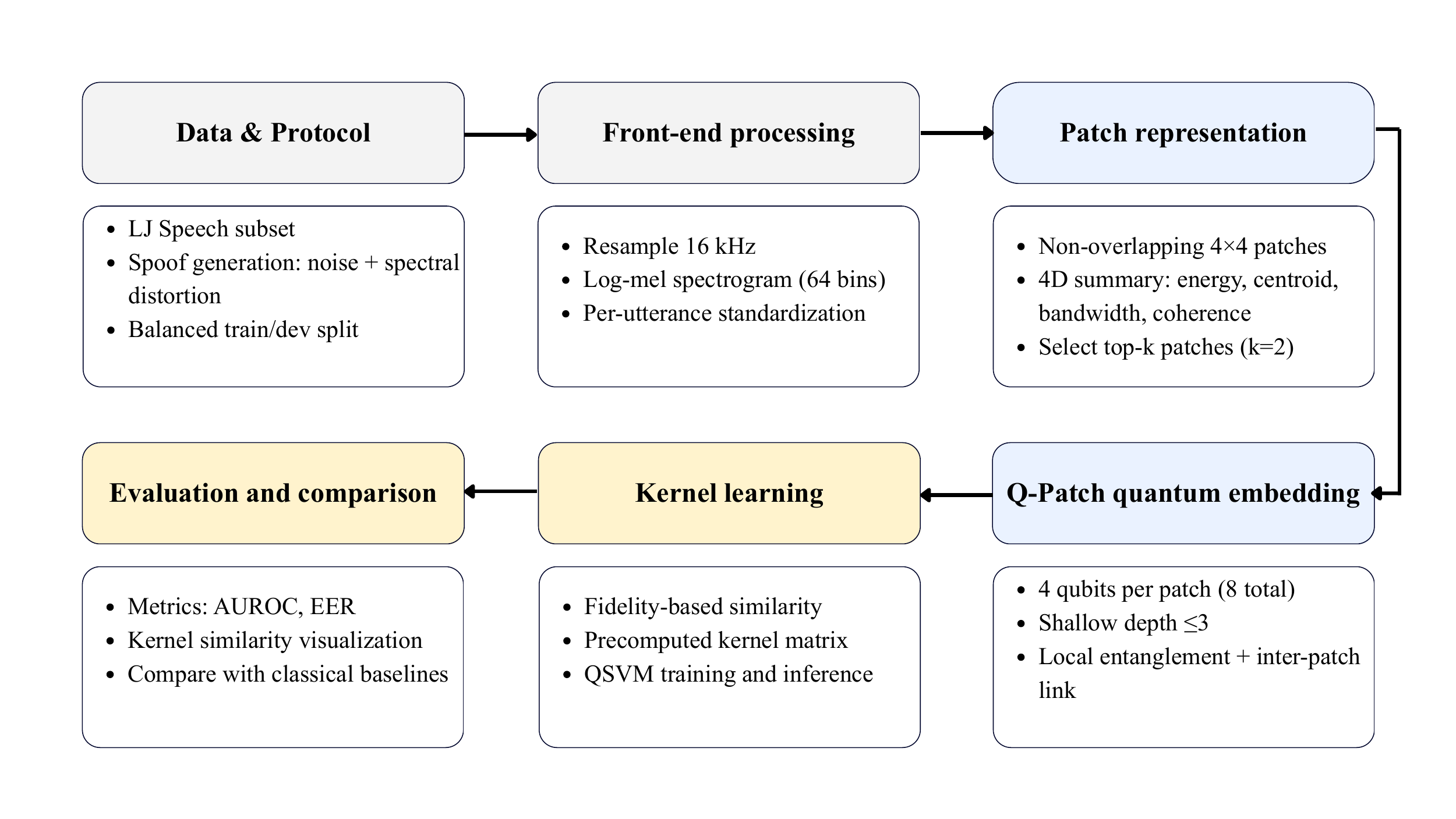}
    \caption{Overview of the Q-Patch pipeline from data construction and time--frequency patch summarization to quantum embedding, kernel learning with QSVM, and evaluation against matched classical baselines.}
    \label{fig:framework}
\end{figure*}

Fig.~\ref{fig:framework} summarizes the end-to-end framework, from data preparation and patch summarization to quantum embedding, kernel learning, and evaluation.

\subsection{Dataset and Experimental Setup}
Due to the computational cost of quantum-kernel simulation, we construct a balanced 100-sample subset from LJ Speech \cite{_17}, consisting of 50 bona fide utterances and 50 spoofed utterances. The spoofed samples are generated from the original recordings using additive Gaussian noise and spectral distortions as a controlled proxy for authenticity manipulation. We then partition this 100-sample set into 80 training and 20 development (dev) samples with preserved class balance and no overlap between splits. This compact protocol supports repeated kernel simulation and controlled comparison with classical baselines, but it should be interpreted as a feasibility study rather than evidence of broad real-world generalization.

\subsection{Front-End Processing}
\begin{figure}
    \centering
    \begin{subfigure}{0.5\textwidth}
        \centering
        \includegraphics[width=\linewidth, height=4cm]{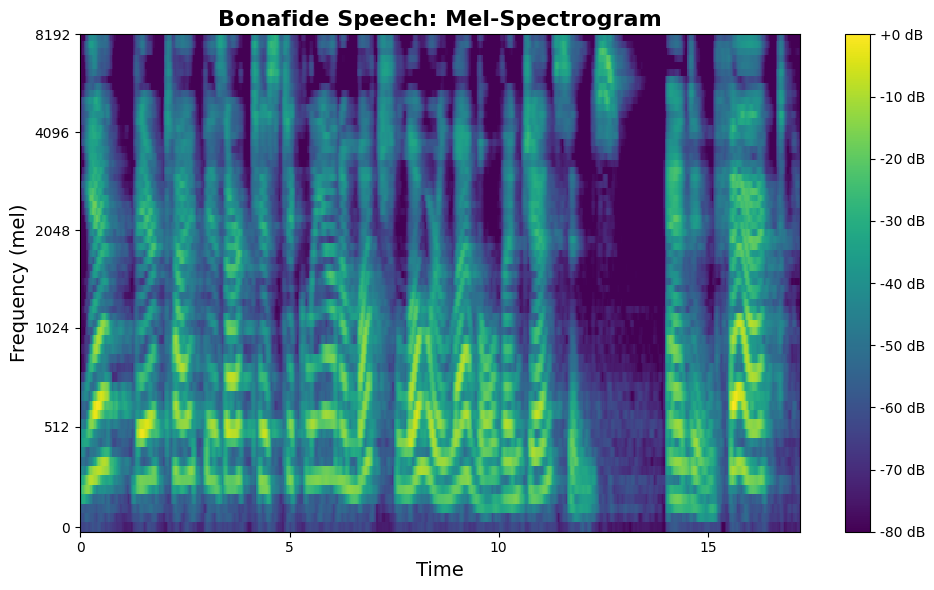}
        \caption{Bona fide}
        \label{fig:bona_fide}
    \end{subfigure}
    \hfill
    \begin{subfigure}{0.5\textwidth}
        \centering
        \includegraphics[width=\linewidth, height=4cm]{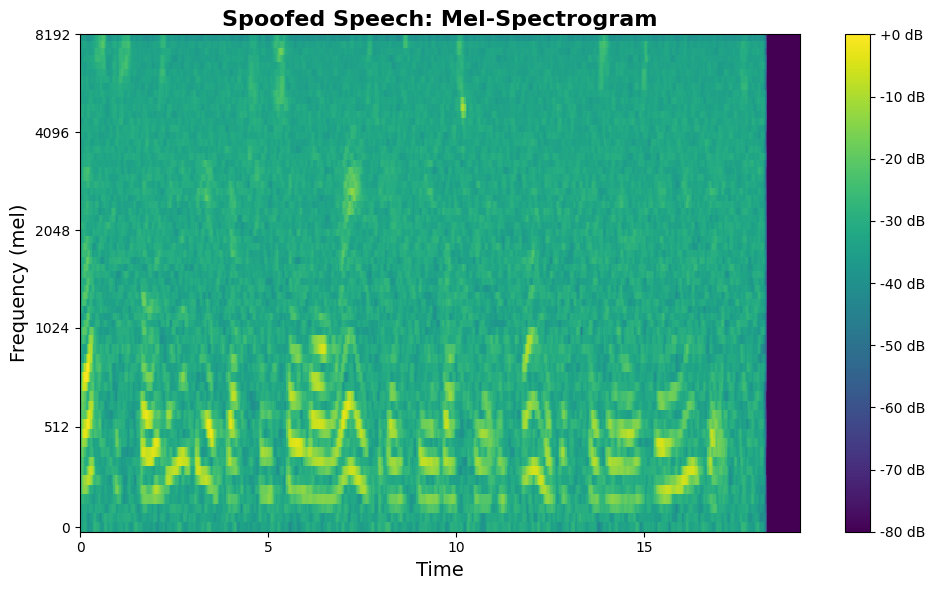}
        \caption{Spoofed}
        \label{fig:spoofed}
    \end{subfigure}
    \caption{Example spectrograms extracted from the LJ Speech dataset: (a) bona fide sample and (b) spoofed sample. Distinct local spectral patterns motivate patch-level modeling for classification.}
    \label{fig:spec_example}
\end{figure}

Each waveform $x[n]$ is resampled to 16~kHz and converted into a log-mel spectrogram $M \in \mathbb{R}^{T \times F}$ with $F = 64$ mel bins. We compute the short-time Fourier transform (STFT) using a 25~ms Hann window and a 10~ms hop, with a 1024-point fast Fourier transform (FFT):
\begin{equation}
X(k,\tau)=\sum_{n} x[n]\;w[n-\tau]\;e^{-j2\pi kn/N},
\end{equation}
where $w[\cdot]$ is the Hann window and $N=1024$. Mel filterbank energies are then obtained as
\begin{equation}
E(f,\tau)=\sum_{k} |X(k,\tau)|^2\,H_f(k),
\end{equation}
where $H_f(\cdot)$ denotes the $f$-th mel filter. Log compression yields
\begin{equation}
M(\tau,f)=\log\left(E(f,\tau)+\epsilon\right),
\end{equation}
with $\epsilon>0$ for numerical stability. To reduce utterance-level scale variation, we apply per-utterance standardization:
\begin{equation}
\tilde{M}(\tau,f)=\frac{M(\tau,f)-\mu_M}{\sigma_M+\epsilon},
\end{equation}
where $\mu_M$ and $\sigma_M$ are computed over all $(\tau,f)$ locations in the utterance. Fig.~\ref{fig:spec_example} shows representative bona fide and spoofed spectrograms after preprocessing.

\subsection{Patch Partitioning and Summary Vectors}
We partition the standardized spectrogram $\tilde{M} \in \mathbb{R}^{T \times F}$ into non-overlapping $4 \times 4$ time--frequency patches. Let $\mathcal{P}=\{P_i\}_{i=1}^{N_p}$ denote the patch set, with
\begin{equation}
P_i=\tilde{M}[\tau_i:\tau_i+3,\; f_i:f_i+3]\in\mathbb{R}^{4\times 4}.
\end{equation}
Each patch $P$ is mapped to a four-dimensional summary vector $\mathbf{s}(P)=[s_1,s_2,s_3,s_4]^\top$ that captures local intensity distribution and short-term temporal consistency:
{\small
\begin{align}
s_1(P) &= \frac{1}{|T||F|} \sum_{\tau \in T} \sum_{f \in F} P_{\tau,f}
\quad \text{(Mean Patch Activation)}, \label{eq:s1}\\[2pt]
\bar{m}_f(P) &= \frac{1}{|T|}\sum_{\tau\in T} P_{\tau,f}, \label{eq:mf}\\[2pt]
w_f(P) &= \frac{(|\bar{m}_f(P)|+\epsilon)}{\sum_{f'\in F} (|\bar{m}_{f'}(P)|+\epsilon)}, \label{eq:wf}\\[2pt]
s_2(P) &= \sum_{f\in F} f\,w_f(P)
\quad \text{(Spectral Centroid)}, \label{eq:s2}\\[2pt]
s_3(P) &= \sqrt{\sum_{f\in F} (f-s_2(P))^2\,w_f(P)}
\quad \text{(Spectral Bandwidth)}, \label{eq:s3}\\[2pt]
\begin{split}
s_4(P) &= \frac{1}{|T|-1}\sum_{\tau}\frac{\langle P_{\tau,:},P_{\tau+1,:}\rangle}
{\|P_{\tau,:}\|_2\,\|P_{\tau+1,:}\|_2+\epsilon} \\
&\quad \text{(Inter-frame Coherence)}. \label{eq:s4}
\end{split}
\end{align}
}
Here $T$ indexes the four patch frames and $F$ indexes the four patch frequency bins. We use $|\bar{m}_f(P)|$ in \eqref{eq:wf} to ensure nonnegative weights under per-utterance standardization.

\paragraph{Top-$k$ patch selection.}
To focus quantum resources on the most informative regions, we rank patches using a simple energy proxy and retain the top two. In our implementation, the ranking score is the mean activation $s_1(P_i)$:
\begin{equation}
\mathcal{I}_2=\mathrm{Top2}\left(\{s_1(P_i)\}_{i=1}^{N_p}\right).
\end{equation}
We use $s_1(P_i)$ as a deterministic salience proxy that favors locally prominent time--frequency regions while keeping the selection rule training-free and computationally inexpensive under a strict qubit budget. We do not claim that mean activation is universally optimal; it may underweight quieter but discriminative regions, and sensitivity to alternative criteria, patch sizes, or small perturbations in spoof generation remains an important direction for future work. Under fixed preprocessing, however, the top-$k$ selection is deterministic for each utterance.

The resulting feature vector is the concatenation of two patch summaries:
\begin{equation}
\mathbf{x}=[\mathbf{s}(P_{i_1});\mathbf{s}(P_{i_2})]\in\mathbb{R}^{8},\quad \{i_1,i_2\}=\mathcal{I}_2,
\end{equation}
which maps naturally to an eight-qubit system, with four qubits assigned to each selected patch. Algorithm~\ref{alg:qpatch_extract} summarizes the full feature-extraction and patch-selection procedure.

\begin{algorithm}[t]
\caption{Q-Patch feature extraction and top-$k$ patch selection}
\label{alg:qpatch_extract}
\begin{algorithmic}[1]
\Require Audio waveform $x[n]$, sampling rate $16$ kHz, mel bins $F=64$, FFT size $N=1024$, window $25$ ms, hop $10$ ms, patch size $(h,w)=(4,4)$, top-$k=2$
\Ensure Concatenated feature vector $\mathbf{x}\in\mathbb{R}^{4k}$
\State Compute log-mel spectrogram $M\in\mathbb{R}^{T\times F}$ from $x[n]$
\State Standardize per utterance: $\tilde{M} \gets (M-\mu_M)/(\sigma_M+\epsilon)$
\State Partition $\tilde{M}$ into non-overlapping patches $\mathcal{P}=\{P_i\}_{i=1}^{N_p}$, each $P_i\in\mathbb{R}^{h\times w}$
\For{$i = 1$ to $N_p$}
    \State Compute $\mathbf{s}(P_i)=[s_1,s_2,s_3,s_4]^\top$ using Eqs.~\eqref{eq:s1}--\eqref{eq:s4}
    \State Assign patch score $e_i \gets s_1(P_i)$
\EndFor
\State Select indices $\mathcal{I}_k \gets \mathrm{TopK}(\{e_i\}_{i=1}^{N_p}, k)$
\State Form $\mathbf{x} \gets [\mathbf{s}(P_{i_1});\dots;\mathbf{s}(P_{i_k})]$ where $\mathcal{I}_k=\{i_1,\dots,i_k\}$
\State \Return $\mathbf{x}$
\end{algorithmic}
\end{algorithm}

\subsection{Quantum Feature Map (Q-Patch)}
\label{subsec:qpatch}
Given a patch summary vector $\mathbf{s}=[s_1,s_2,s_3,s_4]^\top$, Q-Patch embeds $\mathbf{s}$ into a four-qubit quantum state using parameterized single-qubit rotations followed by a lightweight entangling chain. Let $U_\phi(\mathbf{s})$ denote the embedding unitary:
\begin{align}
U_{\mathrm{rot}}(\mathbf{s}) &= R_X(q_0, s_1)\, R_Y(q_1, s_2)\, R_Z(q_2, s_3)\, R_Y(q_3, s_4), \label{eq:urot}\\
U_{\mathrm{ent}} &= CZ(q_0,q_1)\,CZ(q_1,q_2)\,CZ(q_2,q_3), \label{eq:uent}\\
U_\phi(\mathbf{s}) &= U_{\mathrm{ent}}\,U_{\mathrm{rot}}(\mathbf{s}), \\
|\phi(\mathbf{s})\rangle &= U_\phi(\mathbf{s})\,|0\rangle^{\otimes 4}. \label{eq:phi}
\end{align}

\begin{figure}[t]
\centering
\footnotesize
\setlength{\fboxsep}{1.2pt}
\setlength{\fboxrule}{0.35pt}
\begin{quantikz}[row sep=0.35cm, column sep=0.28cm]
\lstick{$q_0$} & \rotgate{R_X(s^{(1)}_1)} & \ctrl{1} & \qw      & \qw            & \qw            & \qw \\
\lstick{$q_1$} & \rotgate{R_Y(s^{(1)}_2)} & \entgate{Z} & \ctrl{1} & \qw       & \qw            & \qw \\
\lstick{$q_2$} & \rotgate{R_Z(s^{(1)}_3)} & \qw      & \entgate{Z} & \ctrl{1} & \qw            & \qw \\
\lstick{$q_3$} & \rotgate{R_Y(s^{(1)}_4)} & \qw      & \qw      & \entgate{Z}     & \ctrl{1} & \qw \\
\lstick{$q_4$} & \rotgate{R_X(s^{(2)}_1)} & \ctrl{1} & \qw      & \qw            & \linkgate{Z}   & \qw \\
\lstick{$q_5$} & \rotgate{R_Y(s^{(2)}_2)} & \entgate{Z} & \ctrl{1} & \qw       & \qw            & \qw \\
\lstick{$q_6$} & \rotgate{R_Z(s^{(2)}_3)} & \qw      & \entgate{Z} & \ctrl{1} & \qw            & \qw \\
\lstick{$q_7$} & \rotgate{R_Y(s^{(2)}_4)} & \qw      & \qw      & \entgate{Z}     & \qw      & \qw \\
\end{quantikz}
\caption{Q-Patch feature map for two selected patches (8 qubits). Rotation encodings are shown in blue, controlled-$Z$ targets in gray, and the inter-patch connection between $q_3$ and $q_4$ is highlighted in amber. The block can be repeated for depth $d \le 3$.}
\label{fig:circuit}
\end{figure}
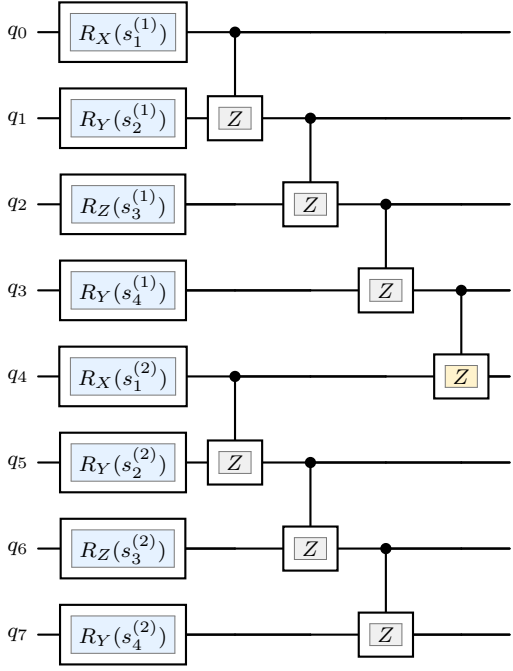

\paragraph{Two-patch encoding (8 qubits).}
For the top two patches, we apply the same embedding to two disjoint four-qubit blocks in parallel. Let $\mathbf{s}^{(1)}$ and $\mathbf{s}^{(2)}$ denote the selected patch summaries. The joint embedding is
\begin{equation}
|\phi(\mathbf{s}^{(1)},\mathbf{s}^{(2)})\rangle
=
\left(CZ(q_3,q_4)\right)
\left(U_\phi(\mathbf{s}^{(1)})\otimes U_\phi(\mathbf{s}^{(2)})\right)
|0\rangle^{\otimes 8},
\end{equation}
where $CZ(q_3,q_4)$ introduces a single inter-patch interaction consistent with spatial adjacency. To remain compatible with Noisy Intermediate-Scale Quantum (NISQ) constraints \cite{_15}, circuit depth is limited to $d \leq 3$. Single-qubit rotations are applied in parallel, and entangling operations are restricted to local controlled-$Z$ chains, as illustrated in Fig.~\ref{fig:circuit}.

In all reported experiments, the two four-dimensional patch summaries in the concatenated feature vector are used directly as rotation angles in the embedding circuit; no additional learned rescaling or variational parameters are introduced. This preserves a controlled comparison with the RBF-SVM, since both methods operate on the same compact patch descriptor.

\paragraph{Kernel definition.}
We define the quantum kernel between two inputs $\mathbf{x}$ and $\mathbf{x}'$ through state fidelity:
\begin{equation}
\kappa(\mathbf{x},\mathbf{x}') =
\left|\left\langle \phi(\mathbf{x}) \,\middle|\, \phi(\mathbf{x}') \right\rangle\right|^2,
\label{eq:qkernel}
\end{equation}
where $|\phi(\mathbf{x})\rangle$ is the embedded state constructed from the selected patch summaries. If more than two patches are used, we process non-overlapping patch pairs and define the final similarity as the arithmetic mean of the corresponding per-pair fidelities. Because nonnegative sums and averages of valid kernels remain positive semi-definite, this construction preserves QSVM compatibility while allowing scalability under bounded depth.

\subsection{QSVM Training and Inference}
Given a training set $\{(\mathbf{x}_i,y_i)\}_{i=1}^{N}$ with labels $y_i \in \{-1,+1\}$, QSVM training uses the precomputed Gram matrix $K \in \mathbb{R}^{N \times N}$:
\begin{equation}
K_{ij}=\kappa(\mathbf{x}_i,\mathbf{x}_j),
\end{equation}
where $\kappa(\cdot,\cdot)$ is defined in \eqref{eq:qkernel}. At inference time, for a test example $\mathbf{x}^\ast$, we compute kernel values against the training set:
\begin{equation}
K^\ast_j=\kappa(\mathbf{x}^\ast,\mathbf{x}_j),\quad j=1,\dots,N,
\end{equation}
which corresponds to a standard kernel support vector machine decision function evaluated in the implicit feature space induced by the quantum embedding \cite{_11,_12}. Algorithm~\ref{alg:qsvm_kernel} details this procedure.

Unlike variational quantum classifiers, Q-Patch does not require iterative optimization of circuit parameters. Once the Gram matrix is computed, learning reduces to the standard convex support vector machine optimization with a precomputed kernel, so convergence is inherited from the underlying solver rather than from a separate non-convex quantum training loop.

\begin{algorithm}[t]
\caption{Quantum kernel construction and QSVM training/inference}
\label{alg:qsvm_kernel}
\begin{algorithmic}[1]
\Require Training set $\{(\mathbf{x}_i,y_i)\}_{i=1}^{N}$ with $y_i\in\{-1,+1\}$, test set $\{\mathbf{x}^\ast_u\}_{u=1}^{N^\ast}$, embedding circuit $U_\phi(\cdot)$ with depth $d\leq 3$
\Ensure QSVM decision scores for test samples
\Function{Kernel}{$\mathbf{x},\mathbf{x}'$}
    \State Prepare $|\phi(\mathbf{x})\rangle \gets U_\phi(\mathbf{x})|0\rangle^{\otimes n}$
    \State Prepare $|\phi(\mathbf{x}')\rangle \gets U_\phi(\mathbf{x}')|0\rangle^{\otimes n}$
    \State \Return $\kappa(\mathbf{x},\mathbf{x}') \gets \left|\langle \phi(\mathbf{x}) \mid \phi(\mathbf{x}') \rangle\right|^2$
\EndFunction
\State \textbf{Training kernel:} initialize $K\in\mathbb{R}^{N\times N}$
\For{$i=1$ to $N$}
    \For{$j=1$ to $N$}
        \State $K_{ij}\gets$ \Call{Kernel}{$\mathbf{x}_i,\mathbf{x}_j$}
    \EndFor
\EndFor
\State Train QSVM with precomputed kernel matrix $K$ \cite{_11,_12}
\State \textbf{Test kernel:} initialize $K^\ast\in\mathbb{R}^{N^\ast\times N}$
\For{$u=1$ to $N^\ast$}
    \For{$j=1$ to $N$}
        \State $K^\ast_{uj}\gets$ \Call{Kernel}{$\mathbf{x}^\ast_u,\mathbf{x}_j$}
    \EndFor
\EndFor
\State Compute QSVM decision scores using $K^\ast$
\end{algorithmic}
\end{algorithm}

\subsection{Comparative Baselines}
We evaluate Q-Patch against two baselines representing complementary classical modeling strategies.

\paragraph{RBF-SVM}
A classical support vector machine with a radial basis function kernel is trained on the same patch-derived input $\mathbf{x}\in\mathbb{R}^{8}$, obtained by concatenating the top-$k$ summary vectors. This baseline isolates the effect of the quantum feature map while holding the input representation fixed.

\paragraph{Tiny CNN}
A compact convolutional neural network with $\leq$100k parameters is trained on spectrogram inputs. This baseline captures local time--frequency patterns without explicit patch summarization.

\paragraph{Evaluation metrics}
We report the area under the receiver operating characteristic curve (AUROC) and equal error rate (EER). AUROC summarizes ranking quality across all thresholds, while EER is standard in anti-spoofing and directly characterizes the false-accept/false-reject trade-off at a single operating point. EER is computed at the operating point $\tau^\star$ where the false positive rate (FPR) equals the false negative rate (FNR):
\begin{equation}
\mathrm{EER}=\mathrm{FPR}(\tau^\star)
\quad \text{such that}\quad
\mathrm{FPR}(\tau^\star)=\mathrm{FNR}(\tau^\star).
\end{equation}
In addition, we analyze quantum-kernel similarity patterns to assess whether bona fide and spoofed samples form separable structure in the induced kernel space, with particular attention to feasibility under shallow depth constraints and limited-data conditions.

\subsection{Implementation Notes}
Circuit depth is constrained to $d\leq 3$ following NISQ feasibility considerations \cite{_15}. Single-qubit rotations are applied in parallel and entanglement is restricted to local controlled-$Z$ chains, with one additional inter-patch connection for the two-patch encoding, to minimize depth while capturing patch-level correlations. Kernel construction scales quadratically in the number of training examples because it requires pairwise kernel evaluations, which motivates the controlled dataset size used in this study.

All experiments were executed in simulation on a CPU-only environment. The implementation used Python, standard numerical libraries, scikit-learn for support vector machine-based classification, and a quantum simulation framework for kernel-fidelity computation. No physical quantum-hardware runs were performed in this study; kernel fidelities were computed under ideal simulation. The shallow eight-qubit, depth-constrained design was chosen to remain compatible with future execution on near-term quantum devices.

\section{Results and Discussion}
\label{sec:results}

This section presents the evaluation protocol, analyzes the structure induced by the proposed quantum kernel, compares Q-Patch with matched classical baselines, and discusses the practical implications and limitations of the reported results.

\subsection{Evaluation Protocol}
We evaluate Q-Patch on a balanced subset derived from LJ Speech \cite{_17}, using 80 training samples and 20 development (dev) samples drawn from a 100-sample set containing 50 bona fide and 50 spoofed utterances. This intentionally compact setup serves two purposes: it makes repeated quantum-kernel simulation computationally manageable and provides a controlled environment for testing whether patch-based quantum embeddings induce meaningful class structure before moving to larger and more diverse spoofing benchmarks.

Unless stated otherwise, all reported metrics are computed on the dev split using standard decision-score evaluation. We report the area under the receiver operating characteristic curve (AUROC) to summarize ranking quality across thresholds and the equal error rate (EER) to characterize the operating point at which false accepts and false rejects are equal. In addition to these scalar measures, we report kernel-similarity statistics and visualize the kernel matrix to examine how Q-Patch organizes bona fide and spoofed samples in the induced feature space.

Because the dev split contains only 20 samples, the reported AUROC and EER should be interpreted as point estimates from a compact feasibility study rather than as statistically definitive performance claims. Accordingly, we treat the observed gains over the baselines as preliminary evidence and leave repeated-split evaluation, uncertainty estimates, and confidence intervals to future work on larger datasets.

\subsection{Quantum Kernel Structure}
A central premise of Q-Patch is that a shallow, structured quantum feature map applied to informative time--frequency patches can produce a kernel in which bona fide and spoofed utterances occupy more separable regions. We examine this hypothesis through fidelity-based kernel similarities across within-class and cross-class comparisons.

The results in Table~\ref{tab:kernel} show three clear patterns. First, self-similarities for identical patches reach $1.000 \pm 0.000$, which is consistent with the fidelity definition and confirms numerical stability of the embedding and overlap computation. Second, within-class similarities across different patches are lower than self-similarities, indicating that the feature map does not collapse diverse samples into a single representation. Third, and most importantly, cross-class similarities between bona fide and spoofed samples are lower ($0.614$--$0.616$) than same-class similarities across different patches, suggesting that the induced kernel captures class-relevant structure rather than only generic patch variability.

\begin{figure}
\centering
\includegraphics[width=\linewidth]{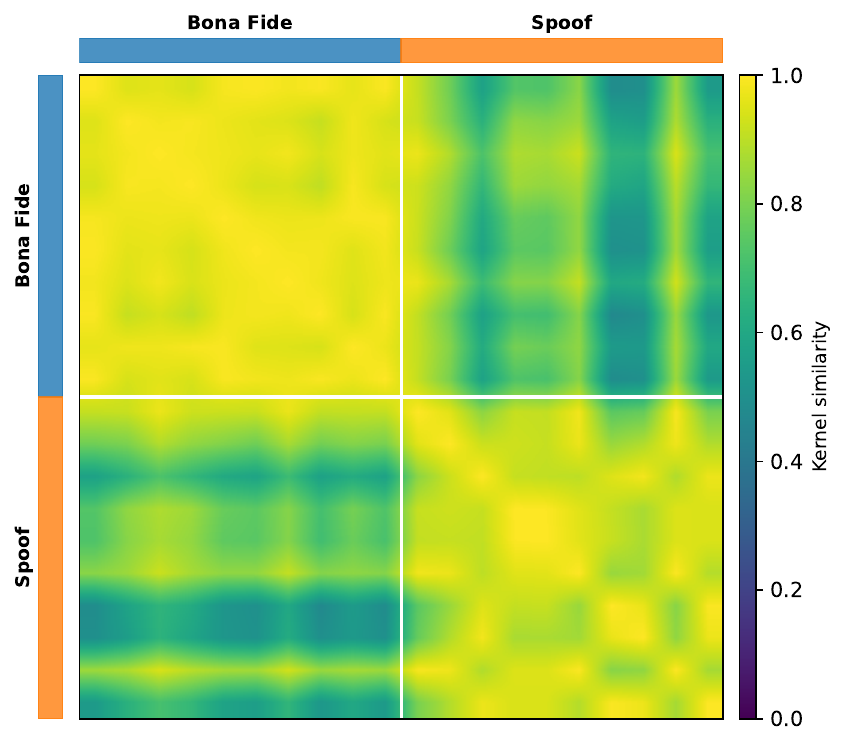}
\caption{Quantum kernel similarity matrix on the development set, with samples ordered by class. Brighter within-class blocks and comparatively darker cross-class regions indicate that Q-Patch induces a class-consistent similarity structure between bona fide and spoofed audio samples.}
\label{fig:kernel_sim}
\end{figure}

In Table~\ref{tab:kernel}, ``same patch'' refers to self-similarity of an identical patch under the fidelity kernel, whereas ``different'' refers to patches drawn from different utterances within the same class. A useful way to interpret the table is to compare the drop in similarity caused by changing the patch within a class with the drop caused by changing the class label itself. The $38.4$--$38.6\%$ reduction observed for bona fide--spoof comparisons matches or exceeds the reduction observed for within-class comparisons across different patches ($32.5$--$37.8\%$). This indicates that the kernel is responsive to authenticity-related cues beyond ordinary within-class variation. The similarity matrix in Fig.~\ref{fig:kernel_sim} reinforces this conclusion by showing visible clustering consistent with class separation in the induced feature space.

\begin{table}
\centering
\small
\setlength{\tabcolsep}{5pt}
\caption{Quantum kernel similarity results on the validation dataset}
\label{tab:kernel}
\begin{tabular}{@{}lcc@{}}
\toprule
\textbf{Comparison Type} & \textbf{Similarity Score} & $\mathbf{\Delta}$ \textbf{from Ideal} \\
\midrule
Bona fide--Bona fide (same patch) & $1.000 \pm 0.000$ & Baseline \\
Spoof--Spoof (same patch)         & $1.000 \pm 0.000$ & Baseline \\
Bona fide--Bona fide (different)  & $0.675 \pm 0.023$ & $-32.5\%$ \\
Spoof--Spoof (different)          & $0.622 \pm 0.018$ & $-37.8\%$ \\
Bona fide--Spoof (patch 1)        & $0.614 \pm 0.015$ & $-38.6\%$ \\
Bona fide--Spoof (patch 2)        & $0.616 \pm 0.017$ & $-38.4\%$ \\
\bottomrule
\end{tabular}
\end{table}

\subsection{Comparative Performance}
Table~\ref{tab:performance} summarizes the classification performance of Q-Patch and the two matched classical baselines. Q-Patch achieves an AUROC of $0.87$ and an EER of $14.8\%$, outperforming both the radial basis function support vector machine (RBF-SVM) trained on the same patch-level descriptors and the compact convolutional neural network (Tiny CNN) trained directly on spectrograms.

The comparison with the RBF-SVM is particularly informative because both methods operate on the same summarized patch features. This makes the RBF-SVM a direct control for assessing the contribution of the proposed quantum feature map and fidelity-based kernel. The improvement over the Tiny CNN further suggests that Q-Patch remains competitive even against a nonlinear model trained on higher-dimensional spectrogram inputs, while maintaining a compact effective quantum footprint of eight qubits and a depth-constrained circuit.

\begin{table}
\centering
\small
\setlength{\tabcolsep}{5pt}
\caption{Performance comparison on the validation dataset}
\label{tab:performance}
\begin{tabular}{@{}lccc@{}}
\toprule
\textbf{Method} & \textbf{AUROC} & \textbf{EER (\%)} & \textbf{Complexity} \\
\midrule
RBF-SVM  & 0.82 & 18.2 & Support vectors \\
Tiny CNN & 0.85 & 16.3 & 98.4K params \\
Q-Patch  & \textbf{0.87} & \textbf{14.8} & \textbf{8 qubits} \\
\bottomrule
\end{tabular}
\end{table}

From an application perspective, the lower EER indicates a more favorable trade-off between false accepts and false rejects under the same evaluation protocol. Taken together with the kernel-space analysis, these results support the view that the proposed patch-based quantum embedding is not only expressive, but also practically discriminative in this controlled setting.

\subsection{Interpretation in the Limited-Data Setting}
Although the proposed protocol uses only 80 training samples, it still represents a low-data regime relative to typical deep anti-spoofing pipelines. Q-Patch is well suited to such conditions because it combines two desirable properties: aggressive dimensionality reduction through patch summarization and top-$k$ selection, and margin-based learning in kernel space.

Concretely, each patch summary $\mathbf{s}\in\mathbb{R}^{4}$ is embedded into a quantum state in $\mathcal{H}=\mathbb{C}^{2^n}$, with $n=4$ qubits per patch block, and similarity is measured through the fidelity kernel
\begin{equation}
k(\mathbf{x},\mathbf{x}')=\left|\langle \phi(\mathbf{x}) \mid \phi(\mathbf{x}')\rangle\right|^2,
\end{equation}
where $\phi(\cdot)$ denotes the implicit feature map induced by the Q-Patch circuit. This construction introduces controlled nonlinearity through shallow entanglement while avoiding the large number of trainable parameters typically associated with deep models. In practical terms, the resulting Quantum Support Vector Machine (QSVM) promotes large-margin separation in the induced feature space, which can help reduce overfitting when labeled data are limited.

\subsection{Practical Feasibility and Scalability}
Q-Patch is designed with near-term quantum constraints in mind. The two-patch configuration requires an eight-qubit circuit with depth limited to $d \leq 3$ and a minimal entangling pattern consisting of three controlled-$Z$ gates per patch block plus a single inter-patch connection, which is consistent with Noisy Intermediate-Scale Quantum (NISQ) feasibility guidelines \cite{_15}. Because single-qubit rotations can be executed in parallel, the circuit remains shallow and comparatively lightweight.

At the same time, the main computational bottleneck lies in kernel construction. For a training set of size $N$, QSVM requires $\mathcal{O}(N^2)$ kernel evaluations. In simulation, these values are obtained through state overlaps; on hardware, they would need to be estimated through repeated measurements with a shot budget $M$, creating a trade-off among dataset size, estimation variance, and runtime. This is precisely where the patch-summarization strategy becomes valuable: it limits qubit count and circuit depth while preserving informative structure from the time--frequency representation.

\subsection{Limitations and Scope}
The present study is intended as a controlled feasibility analysis, and several limitations should be kept in mind when interpreting the results.

\paragraph{Controlled spoof generation.}
The spoofed samples are generated through additive noise and spectral distortions. These manipulations provide a useful controlled proxy for authenticity variation, but they do not capture the full diversity of real-world attacks, including replay conditions, modern neural text-to-speech systems, and voice conversion artifacts. The reported results should therefore be interpreted as evidence of discriminative potential in a controlled setting rather than as a claim of comprehensive anti-spoof robustness.

\paragraph{Dataset size and statistical uncertainty.}
The dev split contains only 20 samples, which limits the granularity of performance estimation and increases uncertainty in the reported AUROC and EER values. Although the kernel-space trends are consistent, broader evaluation protocols will be needed to establish stronger claims about generalization.

\paragraph{Simulation-to-hardware gap.}
The current study evaluates kernel values under ideal quantum simulation. Real quantum hardware introduces noise, gate errors, decoherence, and finite-shot effects, all of which can perturb fidelity estimates. Although Q-Patch is deliberately designed to remain shallow and hardware-aware \cite{_15}, additional studies under realistic noise models or on physical devices will be necessary to characterize deployment robustness. For this reason, the present results should be regarded as ideal-simulation feasibility results rather than hardware-level performance estimates.

\paragraph{Sensitivity to preprocessing and patch selection.}
The proposed framework depends on the preprocessing pipeline, the patch size, the top-$k$ criterion, and the selected summary statistics, all of which influence which spectrogram regions are ultimately encoded. A broader ablation study over patch size, $k$, circuit depth, and entanglement pattern was not feasible within the present compute and page budget. The reported configuration should therefore be viewed as a principled design point rather than an exhaustive exploration of the design space.

\subsection{Sustainability Considerations}
No direct emissions estimate was made for the present study. Since all results are based on small-scale simulation, carbon reporting is best framed here as a recommendation for future larger-scale or hardware-based evaluations rather than as a completed measurement. A practical next step would be to record runtime, energy consumption, and estimated CO$_2$e during feature extraction, model training, and kernel computation using standard accounting tools. Such reporting would support more transparent comparisons of accuracy--compute trade-offs between classical baselines and quantum-kernel approaches as evaluation scales increase.

\section{Conclusion}
\label{sec:conclusion}

This paper presented \textit{Q-Patch}, a patch-based quantum feature-mapping framework for audio spoofing detection that encodes informative local time--frequency regions into shallow quantum circuits and performs classification through a fidelity-based kernel. Experimental results on a controlled and balanced validation setup showed that Q-Patch induces class-consistent kernel structure and improves performance over matched classical baselines, indicating that time--frequency-aware quantum feature maps can provide a useful inductive bias for audio authenticity discrimination in limited-data settings. In addition, the proposed framework remains computationally lightweight by combining a shallow, deterministic quantum embedding with standard kernel-based classification.

The present work should be viewed as a feasibility study under compute and circuit-depth constraints. Future research will focus on evaluation over larger and more diverse spoofing benchmarks, robustness under realistic noise and hardware effects, and scalable kernel-estimation or approximation methods for practical deployment.

\balance

\bibliographystyle{IEEEtran}
\bibliography{Sample-base}

\end{document}